\documentclass{article}

\usepackage[final]{neurips_2023}
\usepackage{amsthm}
\usepackage{amsmath}
\usepackage{graphicx}
\usepackage[utf8]{inputenc} 
\usepackage[T1]{fontenc}    
\usepackage{hyperref}       
\usepackage{url}            
\usepackage{booktabs}       
\usepackage{amsfonts}       
\usepackage{nicefrac}       
\usepackage{microtype}      
\usepackage{xcolor}         
\usepackage{caption}
\usepackage{subcaption}
\usepackage{stmaryrd}
\usepackage{geometry}
\usepackage{float}
\usepackage{siunitx}
\usepackage{tabularray}
\usepackage{colortbl} 

\definecolor{MIGColor}{rgb}{0.8, 0.9, 1.0} 
\definecolor{JEMMIGColor}{rgb}{1.0, 0.9, 0.8} 
\definecolor{IRSColor}{rgb}{1.0, 0.8, 0.8} 
\definecolor{ExplicitnessColor}{rgb}{0.9, 0.8, 1.0} 
\definecolor{LightGrayColor}{rgb}{0.9, 0.9, 0.9} 

\PassOptionsToPackage{numbers}{natbib}

\title{Learning Disentangled Speech Representations}

\author{Yusuf Brima $^{1,2,*}$ \quad Ulf Krumnack$^{1}$ \quad Simone Pika$^{2}$ \quad Gunther Heidemann$^{1}$ \\
$^1$Computer Vision, $^2$Comparative BioCognition, $^*$corresponding author\\
Institute of Cognitive Science, Universität Osnabrück, Germany\\
\texttt{\{ybrima,krumnack,spika,gheidema\}@uos.de}}

\begin{document}

\maketitle

\begin{abstract}
Disentangled representation learning in speech processing has lagged behind other domains, largely due to the lack of datasets with annotated generative factors for robust evaluation. To address this, we propose SynSpeech, a novel large-scale synthetic speech dataset specifically designed to enable research on disentangled speech representations. SynSpeech includes controlled variations in speaker identity, spoken text, and speaking style, with three dataset versions to support experimentation at different levels of complexity.

In this study, we present a comprehensive framework to evaluate disentangled representation learning techniques, applying both linear probing and established supervised disentanglement metrics to assess the modularity, compactness, and informativeness of the representations learned by a state-of-the-art model. Using the RAVE model as a test case, we find that SynSpeech facilitates benchmarking across a range of factors, achieving promising disentanglement of simpler features like gender and speaking style, while highlighting challenges in isolating complex attributes like speaker identity. This benchmark dataset and evaluation framework fills a critical gap, supporting the development of more robust and interpretable speech representation learning methods.
\end{abstract}
\textbf{Keywords:} disentangled representation learning, synthetic speech dataset, speech processing, generative modeling, supervised disentanglement metrics, SynSpeech

\section{Introduction}
\label{sec:intro}
Learning disentangled representations yields promising results in domains like computer vision, allowing models to robustly separate generative factors within data~\cite{locatello2019challenging,higgins2017beta,kim2018disentangling}. A disentangled representation \textit{independently} captures the ``true'' generative factors of variation that explain the data. Such representations provide multiple benefits including \textit{enhanced predictive abilities} on downstream tasks, \textit{decreased sample complexity}, \textit{greater explainability}, \textit{fairness}, and a means to avoid \textit{shortcut learning}~\cite{carbonneau2020measuring}. However, progress on learning disentangled speech representations has been limited, despite potential benefits such as improved understanding of speech signals, interpretable features, controllable generation, source separation, multilingual and cross-lingual speech processing, voice conversion, robustness to variability, few-shot learning, privacy-preserving, etc\cite{Qian2018UnsupervisedSL,chou2018multi}.

A key challenge impeding advancement in this domain is the lack of suitable benchmarking datasets for quantitative evaluation and analysis~\cite{hsu2017unsupervised, mo2019semantic}. Real-world speech data often lack explicit ground truth annotation of generative factors. On the other hand, most synthetic datasets are not large or rich enough to benchmark speech disentanglement models~\cite{wang2018style,zhou2020comparison,jia2018transfer}. This contributes to difficulties in reproducible research and standardized comparison of methods.

SynSpeech\footnote{The dataset can be downloaded from the following \href{https://figshare.com/projects/Neural_Speech_Synthesis_for_Disentangled_Representation_Learning/226971}{here}. An interactive demo is available \href{https://synspeech.github.io/}{here} to explore audio samples.} represents, to the best of our knowledge, the first large-scale synthetic speech dataset designed for benchmarking disentangled speech representation learning for both content and speaker characteristics. With its three distinct dataset versions, SynSpeech enabled a comprehensive analysis of representation generalization through LP and facilitated an empirical evaluation of disentanglement using supervised and unsupervised disentanglement metrics. This benchmark dataset provided fundamental insights, advancing understanding and progress in this previously underexplored area of speech representation.

\section{Methodology}
\label{sec:methods}
\subsection{SynSpeech Dataset}
To enable standardized evaluation, we developed SynSpeech, a large-scale synthetic speech dataset designed for controlled experimentation on key generative factors, including speaker identity, spoken text, gender, and speaking style. SynSpeech contains $184,560$ total utterances generated using advanced neural text-to-speech models, ensuring high-quality, naturalistic outputs. Each utterance is annotated with ground truth factors, facilitating isolated variation of individual attributes for precise benchmarking.

SynSpeech includes controlled variability across speaker identities, text content, and speaking styles. We used the GPT-4 Large Language Model (LLM)~\cite{achiam2023gpt} to generate $500$ diverse sentences, spanning contexts from casual dialogues to scientific explanations. For speaker diversity, we synthesized utterances from $249$ speakers, sourced from LibriSpeech100~\cite{panayotov2015librispeech}, including both male and female voices, to balance speaker identity combinations with content.

\begin{figure}[!ht]
    \centering
    \includegraphics[width=\textwidth]{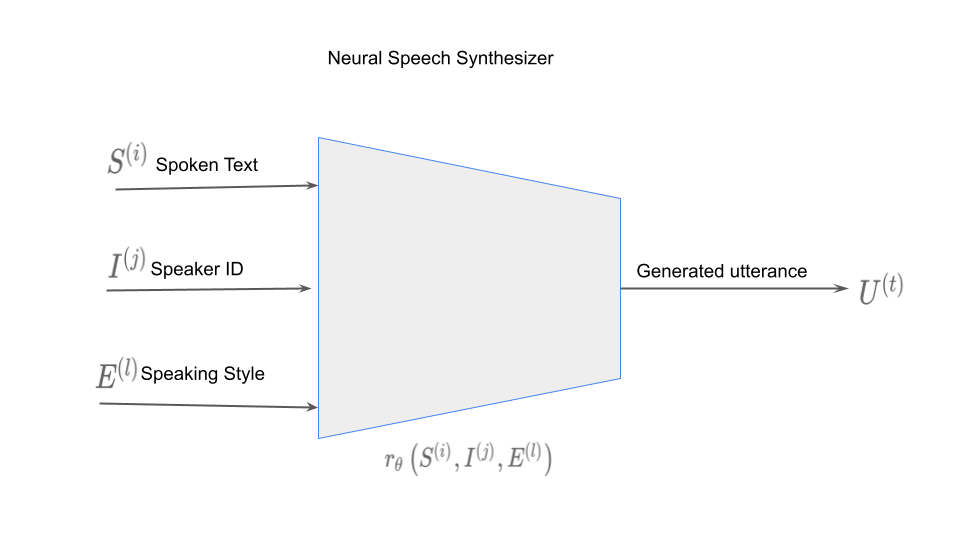}
    \caption{Illustration of the Neural Speech Synthesizer process. The model takes three primary inputs: spoken text $S^{(i)}$, speaker identity $I^{(j)}$, and speaking style $E^{(l)}$. These inputs are combined through the synthesis function $r_\theta(S^{(i)}, I^{(j)}, E^{(l)})$ to generate the final utterance $U^{(t)}$, capturing the specified content, speaker characteristics, and style.}
    \label{fig:Neural_Speech_Synthesizer}
\end{figure}

The generation process is illustrated in Figure~\ref{fig:Neural_Speech_Synthesizer}. The neural speech synthesizer takes three primary inputs—spoken text $S^{(i)}$, speaker identity $I^{(j)}$, and speaking style $E^{(l)}$—and combines them through a synthesis function $r_\theta(S^{(i)}, I^{(j)}, E^{(l)})$ to produce the final utterance $U^{(t)}$, capturing the specified content, speaker characteristics, and style.

Each utterance was synthesized using four speaking styles—\textit{default}, \textit{friendly}, \textit{sad}, and \textit{whispering}—enabled by the OpenVoice toolkit~\cite{qin2023openvoice}, resulting in a 16kHz dataset suited for assessing disentanglement across speaker identity, content, and style.

The dataset is released in three versions, summarized in Table~\ref{tab:synspeech_versions}, to support experimentation at different scales of speaker and content complexity. These versions allow for tailored analyses across diverse configurations of speaker and content diversity.

\begin{table}[H]
    \centering
    \begin{tabular}{@{}lcccc@{}}
        \toprule
        Version & Speakers & Contents & Styles & Total \\ 
        \midrule
        Small  & 50   & 500   & 1 & 25000 \\
        Medium & 25   & 500   & 4 & 50000 \\
        Large  & 249  & 110   & 4 & 109560 \\ 
        \bottomrule
    \end{tabular}
    \caption{Overview of the Dataset Versions}
    \label{tab:synspeech_versions}
\end{table}

\subsection{General Setup}
There is no agreed definition of what constitutes a disentangled representation~\cite{carbonneau2020measuring}. Thus, we first examine what suitable properties are desired in a disentangled representation. A representation has to be \textit{distributed}, meaning, given a dataset $\mathcal{D} = \{\mathbf{x}^{(i)}\}_{i=1}^N$ where each data point $\mathbf{x}^{(i)} \in \mathcal{X}$ is assumed to be a composition of $m$ \textit{disjoint} generative factors from an factor space $\mathcal{V} = \{\mathcal{V}_k \}_{k=1}^m$ through a generative process $g: \mathcal{V} \to \mathcal{X}$ where they fulfill the conditions of \textit{modularity} -- factor must have independent causal influence, \textit{compactness} --  a single dimension ideally encodes the factor in the representation space and \textit{explicitness} -- factors must be useful~\cite{carbonneau2020measuring}. We denote the specific combination of factors used to generate data point $\mathbf{x}^{(i)}$ by $\mathbf{v}^{(i)}$ and denote the collection of all factor realizations used to generate the dataset $\mathcal{D}$ by $\mathcal{D}_V = \{\mathbf{v}^{(i)}\}_{i=1}^N$. 
In such a setting, a linear relation between generative factors and learned latent variables is an ideal trait as it is the most straightforward relation to interpret. However, real-world data is often complex and contains non-linear relations. Gender, for example, is a generative factor that influences voice pitch, but it is not the only factor. Other factors like age, health, and even emotional state can also affect pitch. This non-linearity complicates the process of disentangling gender from voice features.

\begin{figure}[!ht]
    \centering
    \includegraphics[width=\textwidth]{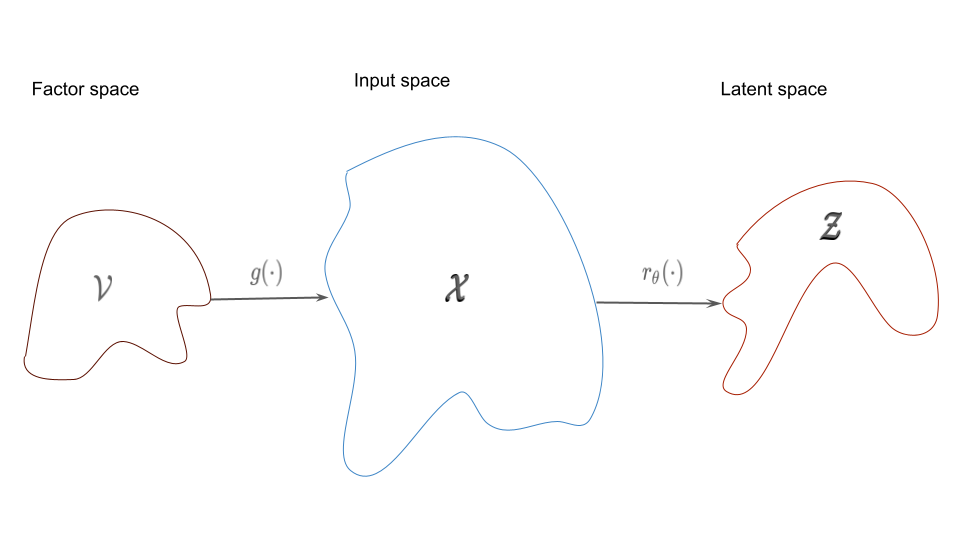}
    \caption{Illustration of supervised disentanglement learning notation. The figure represents the transformation from factor space $\mathcal{V}$, containing generative factors, to input space $\mathcal{X}$ via the generative process $g(\cdot)$. The learned representation function $r_\theta(\cdot)$ further maps input data from $\mathcal{X}$ to latent space $\mathcal{Z}$, where disentangled representations are formed. This setup facilitates the evaluation of disentanglement by comparing the generative factors in $\mathcal{V}$ with their corresponding representations in $\mathcal{Z}$.}
    \label{fig:Disentanglement_Learning}
\end{figure}

Our goal is to learn a function $r_\theta: \mathcal{X} \to \mathcal{Z}$ that maps data points in the input space into a latent space $\mathcal{Z}\in \mathbb{R}^{d}$, with $\mathbf{z}^{(i)}=r_{\theta}(\mathbf{x}^{(i)}) \in \mathcal{Z}$ being the learned representation for the $i$-th data point $\mathbf{x}^{(i)}$.  $\mathcal{D}_Z = \{\mathbf{z}^{(i)}\}_{i=1}^N$ is the set of all data points in $\mathcal{D}$ projected in the learned latent space. So, supervised disentanglement metrics (SDMs) compare the original generating factors $\mathcal{D}_V$ to the learned latent representations $\mathcal{D}_Z$. 
We have illustrated this notation in Figure~\ref{fig:Disentanglement_Learning} and the data generative process in Figure~\ref{fig:Neural_Speech_Synthesizer}.

To evaluate disentanglement on the proposed dataset, we conducted a comprehensive assessment using \textit{linear probing} (LP) and SDMs, following the methodology outlined by~\cite{carbonneau2020measuring}. LP was used to analyze the generalization capacity of learned representations across latent dimensions for various downstream tasks, providing insights into their structure and utility.

Additionally, SDMs were applied to evaluate the extent to which representations captured independent generative factors, offering a rigorous empirical measure of disentanglement. Together, these methods provide a detailed assessment of the model's ability to separate generative factors in synthetic speech data.

\begin{figure}[htbp]
    \centering
    \includegraphics[width=\textwidth]{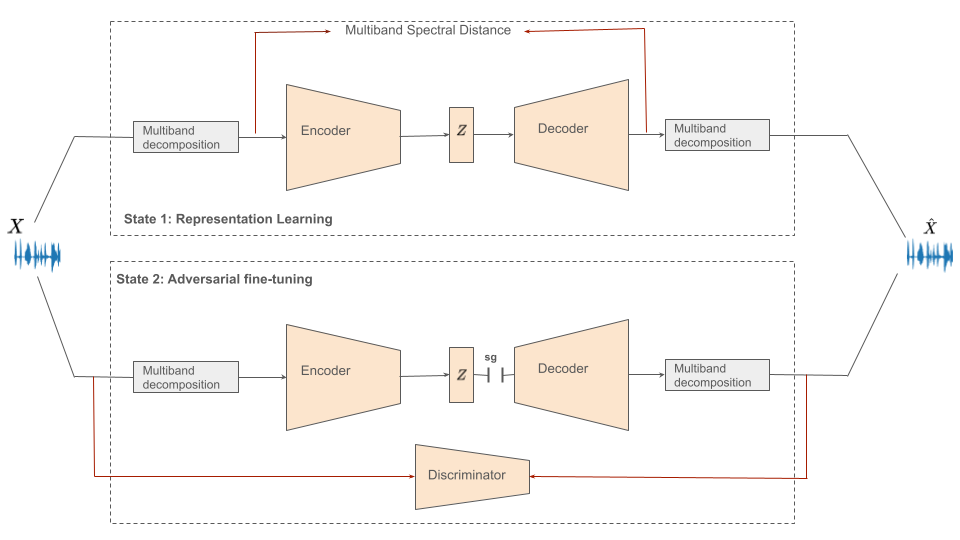}
    \caption{Architecture of the multiband Beta-VAE model with spectral distance and adversarial fine-tuning. This setup includes the multi-band decomposition, which processes each frequency band independently to enhance spectral fidelity, as well as adversarial fine-tuning to improve realism.}
    \label{fig:multiband_beta_vae}
\end{figure}

For our benchmarking study, we adopted the RAVE\footnote{Real-time Audio Variational autoEncoder} methodology~\cite{caillon2021rave} as shown in Figure~\ref{fig:multiband_beta_vae}, which includes multi-band decomposition, a spectral distance objective, adversarial fine-tuning, and singular value decomposition (SVD) to ensure compact and interpretable latent representations. This setup is designed to achieve both high fidelity and realism in audio synthesis, supporting disentanglement analysis in a structured way.

Our evaluation framework thus offers a standardized, quantitative assessment of speech disentanglement performance on the novel dataset. Although synthetic data may not capture the full complexity of real-world audio, this controlled environment establishes a foundational benchmark. If disentanglement methods do not succeed in a setting like this, they are unlikely to do so effectively in more complex, real-world scenarios.

\subsection{Evaluation Metrics}

Following our general setup, we evaluate disentangled representations using a range of supervised metrics. To assess the explicitness of the learned factors and the compactness and modularity of representations, we apply two primary evaluation strategies: Linear Probing and Supervised Disentanglement Evaluation. 

\subsubsection{Linear Probing}

Linear Probing (LP) provides a straightforward approach to gauge the degree to which the latent codes are linearly correlated with known factors of variation. By training a linear classifier or regressor on the latent codes to predict factors, we measure how well the learned representations capture the underlying structure of the data in a linearly accessible way. This approach aligns with the goals of interpretable factor separation in the latent space.

\subsubsection{Supervised Disentanglement Evaluation}

Our supervised disentanglement evaluation follows established metric families to assess modularity, compactness, and explicitness within the learned representations. We categorize these metrics based on their approach to measuring disentanglement—\textit{intervention-based}, \textit{predictor-based}, and \textit{information-based}—drawing on the taxonomy proposed by~\cite{carbonneau2020measuring}. This categorization provides a structured framework for interpreting disentanglement in our results.

\paragraph{Intervention-based Metrics} 
\noindent
Intervention-based metrics evaluate disentanglement by examining variations in the learned representations under controlled conditions where specific factors of variation are held constant while others vary. Our analysis focuses on the Interventional Robustness Score (IRS) as a key metric within this family.

IRS measures how well representations stay stable when only nuisance factors change. IRS works on the idea that varying irrelevant (nuisance) factors should not impact the dimensions representing key (targeted) factors. To calculate IRS, we create a reference set with a fixed target factor, then make a second set where this target factor is the same but nuisance factors differ. The distance between the averages of these sets’ codes, often measured with the $\ell_2$ norm, shows how much the nuisance factors impact the target factor representation. IRS repeats this process, taking the maximum distance each time, and averages these maxima, weighted by how common each target factor is in the data. This approach measures how well target factor representations remain unaffected by irrelevant changes, highlighting the disentanglement quality in our models.

\paragraph{Information-based Metrics}

\noindent

Information-based metrics quantify disentanglement through \textit{mutual information} (MI) between latent dimensions and the ground-truth factors of variation. Mutual information measures the mutual dependency between two random variables, defined as:
\begin{equation}
I(v, z) = \sum_{i=1}^{B_v} \sum_{j=1}^{B_z} P(i, j) \log \left(\frac{P(i, j)}{P(i) P(j)}\right),
\end{equation}
where $B_v$ and $B_z$ are the bins for factor $v$ and latent dimension $z$, respectively, $P(i, j)$ is the joint probability of $i$ and $j$, and $P(i)$ and $P(j)$ are the marginal probabilities. Higher mutual information indicates stronger dependencies, suggesting that the latent dimension $z_j$ effectively captures the variability in factor $v_i$.

\noindent

Using MI as a foundation, we employ two metrics—Mutual Information Gap (MIG) and Joint Entropy Minus Mutual Information Gap (JEMMIG)—to evaluate factor independence, modularity, and compactness within the learned representations.

\noindent

MIG computes the MI between each factor and latent dimension, $I(v_i, z_j)$. For each factor $v_i$, MIG identifies the latent dimension $z_\star$ with the maximum MI, denoted $I(v_i, z_\star)$. It then identifies the second-highest MI for that factor, $I(v_i, z_\circ)$. The difference between these two MI values constitutes the gap, which is then normalized by the entropy of the factor $v_i$:
\begin{equation}
\text{MIG} = \frac{I(v_i, z_\star) - I(v_i, z_\circ)}{\text{H}(v_i)},
\end{equation}
where $\text{H}(v_i)$ is the entropy of the factor $v_i$. The final MIG score is obtained by averaging the normalized gaps across all factors, providing a measure of how well-separated the factors are in the latent space, with higher scores indicating greater disentanglement.

\noindent

Entropy denoted $\text{H}(v_i)$, measures the uncertainty or randomness associated with a factor $v_i$. For a discrete variable $v_i$, entropy is given by:
\begin{equation}
\text{H}(v_i) = - \sum_{k}^{B_v} P(v_i^{k}) \log P(v_i^{k}),
\end{equation}
where $P(v_i^{k})$ is the probability of each possible state $v_i^{k}$ of factor $v_i$. In the context of disentanglement, entropy serves as a normalizing factor, allowing us to interpret mutual information values relative to the overall variability in each factor.
\noindent

While MIG measures the compactness of each factor's representation by focusing on the dimension with maximum MI, it does not account for modularity—ensuring that each latent dimension captures only one factor. To address this, JEMMIG incorporates the joint entropy of the factor and its best corresponding code dimension. For each factor $v_i$, JEMMIG is calculated as:
\begin{equation}
\text{JEMMIG} = \text{H}(v_i, z_\star) - I(v_i, z_\star) + I(v_i, z_\circ),
\end{equation}
where $\text{H}(v_i, z_\star)$ is the joint entropy of factor $v_i$ and the dimension $z_\star$ with the highest MI for $v_i$, and $I(v_i, z_\circ)$ is the MI with the next highest dimension. Unlike MIG, lower JEMMIG scores indicate better disentanglement. The maximum value for JEMMIG is bounded by $\text{H}(v_i) + \log(B_z)$, where $B_z$ is the number of bins used in code space discretization. To normalize JEMMIG to the interval $[0,1]$, we compute:
\begin{equation}
\widehat{\text{JEMMIG}} = 1 - \frac{\text{H}(v_i, z_\star) - I(v_i, z_\star) + I(v_i, z_\circ)}{\text{H}(v_i) + \log(B_z)},
\end{equation}
and report the average $\widehat{\text{JEMMIG}}$ across all factors $v_i$ as the final score. 

\noindent

MIG and JEMMIG offer an information-theoretic perspective on disentanglement, allowing us to evaluate the independence and compactness of factor representations in the latent space. These metrics help identify dimensions with high factor-specific compactness and provide insights into the distribution of factor information across the representation space.

\subsubsection{Predictor-based Metrics}

\noindent

Predictor-based metrics evaluate representations by training classifiers or regressors to predict factor values directly from the latent codes. By analyzing the performance of these predictors, we can assess how well each latent dimension captures specific factors, providing insight into the clarity and interpretability of the representations. This approach is naturally suited to measuring explicitness, as it directly tests the degree to which each code dimension contains information about individual factors.

\paragraph{Explicitness Score}

\noindent

In the present study, we utilize the Explicitness Score, calculated by training a simple classifier, such as logistic regression, on the entire latent code to predict discrete factor values. The classification performance is measured by the area under the ROC curve (AUC-ROC) for each factor, with the final score being the average AUC-ROC across all classes and factors. To normalize the metric to a $[0,1]$ scale, where 0 represents random guessing and 1 indicates perfect classification, we adjust for the AUC-ROC’s minimal value of 0.5. This approach quantifies how well individual latent dimensions explicitly encode specific factors, providing a direct measure of interpretability and aiding in evaluating the model’s ability to achieve clear factor separation.

\noindent

These evaluation methods give us a detailed perspective on disentanglement within our models, encompassing factor independence and interpretability. Using intervention-based, predictor-based, and information-based metrics, we benchmark our models’ ability to isolate and represent underlying factors with minimal entanglement, providing comprehensive insights into the structure of our learned representations.

\section{Results and Analysis}
\label{sec:outcomes}

This work establishes a benchmarking framework to evaluate the effectiveness of the state-of-the-art (SoTA) RAVE model in disentangling key generative factors in speech, including speaker identity, gender, and speaking style. This evaluation focuses on a representation learner's capacity to achieve \textit{modularity}, \textit{compactness}, and \textit{explicitness}—critical characteristics for disentangled representations~\cite{kim2018disentangling,chen2018isolating,kumar2017variational}. By systematically analyzing these attributes, we provide insights into the model's strengths and limitations in capturing distinct generative factors enhancing interpretability and applicability.

Our framework includes two main analyses: \textit{LP of latent dimensions} and \textit{SDMs}. Together, these assessments comprehensively examine how well-trained models isolate and represent sources of variation in speech, offering a detailed perspective on the progress in disentangled speech representation learning.

\subsection{Linear Probing of Latent Dimensions}

\begin{figure}[!htbp]
    \centering
    \includegraphics[width=\textwidth]{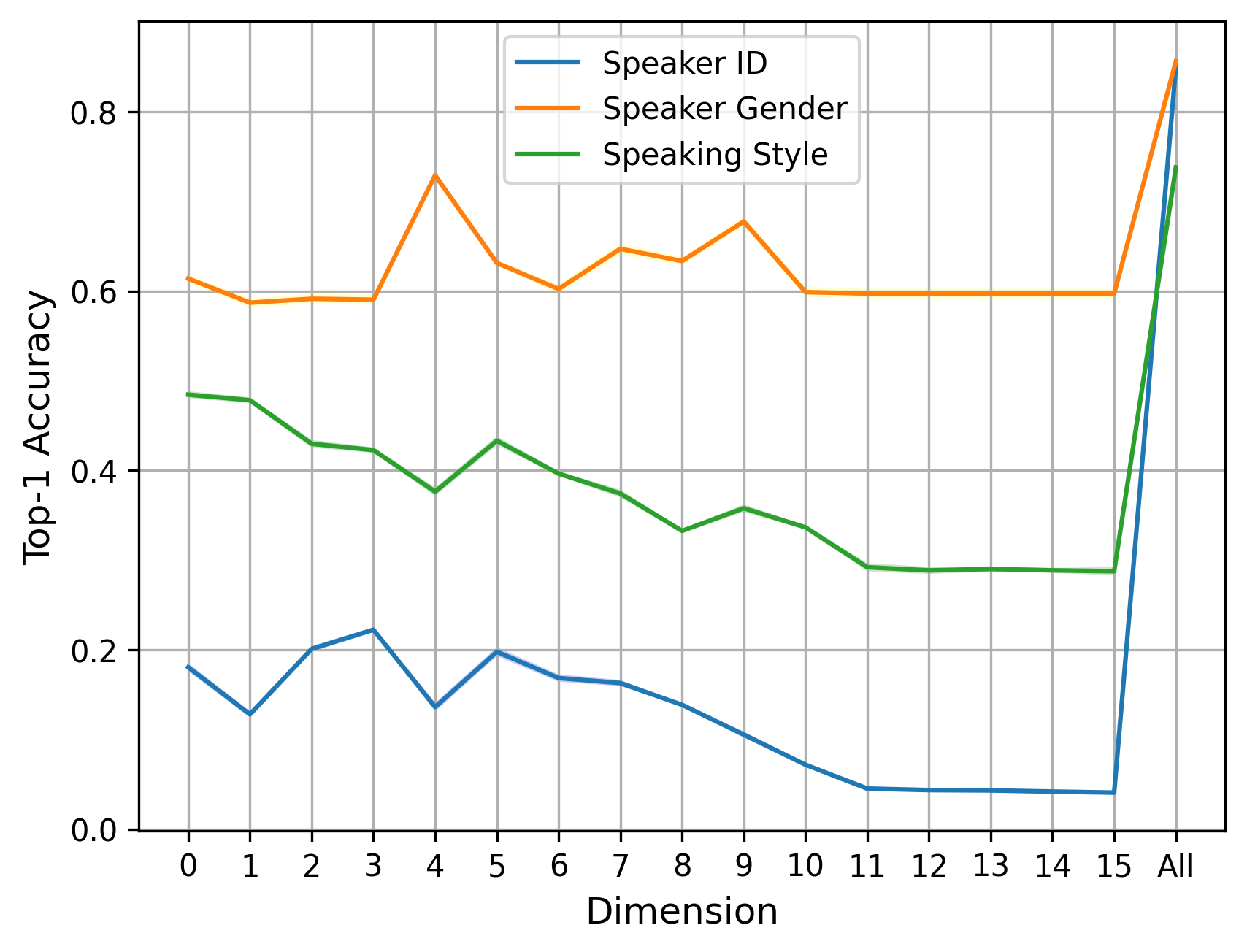}
    \caption{LP Accuracy of Latent Dimensions for Speaker ID, Gender, and Speaking Style on the medium-sized dataset with reported mean and standard deviation for 5 experimental runs.}
    \label{fig:accuracy_trends_humans_medium}
\end{figure}

To evaluate the separability and generalization of latent representations, we applied LP across the model’s latent representations dimension-wise to predict three key attributes: speaker identity (SI), gender, and speaking style. These tasks vary in complexity—speaker gender is a binary classification (male/female), speaking style requires a four-way classification, and speaker identity is a more challenging 25-class problem. Figure~\ref{fig:accuracy_trends_humans_medium} presents the accuracy trends across the latent dimensions for each attribute.

From Figure~\ref{fig:accuracy_trends_humans_medium}, several important patterns emerge. Dimension $4$ stands out with the highest accuracy score for the speaker gender task, indicating that it captures more gender-related information. This suggests the model has learned to encode gender fairly distinctly in this dimension, leveraging specific pitch characteristics that differentiate male and female voices. Such a clear peak for gender accuracy in a single dimension, while there are dips for the two other factors, hints at a degree of successful disentanglement for this attribute. It can also be observed that dimensions $7$ and $9$ relatively encode speaking gender as well.

Dimension $0$ shows the highest test accuracy for speaking style, suggesting that it captures the most concentrated information about this attribute. The stable and moderately high accuracy across other dimensions such as $1$ and $5$ implies that speaking style information may be redundantly encoded. Yet, dimension $1$ specifically holds the most significant style-related information. This indicates partial disentanglement in speaking style, with some degree of distribution across other dimensions.

For speaker identity, the most complex of the tasks, dimension $3$ achieves the highest accuracy, although the accuracy remains relatively low compared to the other tasks. This suggests that while dimension $3$ encodes some identity-specific information, the model struggles to fully disentangle identity due to the inherent complexity of the task and the potential overlap with features related to other factors, such as gender or style. The relatively low accuracy across dimensions for speaker identity reflects this challenge, especially with just a linear classifier with $\mathbb{R}^{256}$ neurons, underscoring the difficulty of isolating identity-related features in a highly entangled latent space.

In an ideally disentangled representation, each attribute would be encoded in a distinct often, orthogonal latent dimension, with minimal overlap in the dimensions that exhibit high accuracy for different tasks. The presence of distinct peaks in dimensions $4$, $0$, and $3$ for gender, speaking style, and speaker identity, respectively, suggests some level of task-specific encoding. However, the lack of clear orthogonality and the moderate accuracy of other dimensions for each task imply that these factors are not fully isolated. This entanglement suggests that while the model partially recovers each factor from the latent codes, it lacks the structural separation needed for complete disentanglement.

The final data point labeled ``All'' shows a significant boost in accuracy for all ground-truth factors, as it aggregates information from all dimensions, resulting in $\mathbb{R}^{16\times 256}$. This increase reflects the added information content available to the classifier rather than effective disentanglement. In this scenario, the latent vectors become entangled, with multiple generative factors coexisting within shared dimensions, enhancing overall accuracy but at the cost of interpretability. Thus, while the ``All'' configuration achieves the best predictive performance, it provides limited insight into the true disentanglement of factors of interest, as it highlights the dependence of factors across dimensions rather than isolating them.

The LP results reveal some task-specific encoding, with dimension $4$ specializing in gender, dimension $0$ in speaking style, and dimension $3$ in speaker identity. However, the overall trends indicate that full orthogonal disentanglement is not achieved, especially for complex attributes like speaker identity. Notably, we observed similar results for the small version of the dataset, as shown in Appendix Figure~\ref{fig:accuracy_trends_humans_small}. These findings highlight the model's ability and potential limitations in creating completely independent representations for each factor of variation, underscoring the need for further refinement in disentangling complex, often multi-dimensional, multi-class attributes in the latent space.
\subsection{Supervised Disentanglement Evaluation}
To assess the model's ability to disentangle key factors, we conducted a supervised evaluation focusing on speaker identity, gender, and speaking style within individual latent dimensions. Using metrics such as MIG, JEMMIG, IRS, and Explicitness, this analysis captures the modularity, compactness, and informativeness of the learned representations—which when combined collectively termed is ``holisticness.'' Specifically, higher MIG and IRS scores indicate stronger holistic disentanglement. In contrast, elevated Explicitness scores suggest clearer, more informative encoding of each ground-truth factor. Conversely, lower JEMMIG values indicate an improvement in disentanglement, highlighting more refined isolation of generative factors across latent dimensions.

\subsubsection{Speaker Identification}

Table~\ref{tab:speaker_id_metrics_overview} summarizes the disentanglement metrics for speaker identification across each latent dimension.
\begin{table}[!ht]
    \centering
    \begin{tabular}{@{}l S[table-format=1.6] S[table-format=1.6] S[table-format=1.6] S[table-format=1.6]@{}}
        \toprule
        Dimension & \multicolumn{2}{c}{\textbf{Information-based}} & \textbf{Intervention-based} & \textbf{Predictor-based} \\ 
        \cmidrule(r){2-3} \cmidrule(r){4-4} \cmidrule(r){5-5}
                  & {MIG $\uparrow$} & {JEMMIG $\downarrow$} & {IRS $\uparrow$} & {Explicitness $\uparrow$} \\ 
        \midrule
        0 & \colorbox{MIGColor}{\num{0.026703392027534967}} & \num{0.22006968321766063} & \num{0.27852030345668694} & \num{0.5812136979017775} \\
        1 & \num{0.017740547730298706} & \num{0.1543613870260635} & \num{0.23790238103877864} & \num{0.4193649454938342} \\
        2 & \num{0.014430466755208468} & \num{0.224464135755424} & \num{0.30279383145590105} & \num{0.5744464814138985} \\
        \rowcolor{LightGrayColor} 
        3 & \num{0.011202663939867546} & \num{0.2210467057016815} & \num{0.28708480612508047} & \colorbox{ExplicitnessColor}{\num{0.5796084259714329}} \\
        4 & \num{0.0006574861208733171} & \num{0.147367404945143} & \num{0.28569738878351875} & \num{0.5240095396697659} \\
        5 & \num{0.005894489712133244} & \num{0.17161612598646714} & \num{0.27230131557263093} & \num{0.526218426133239} \\
        6 & \num{0.001535670154844862} & \num{0.19462207573878765} & \num{0.3030000471526395} & \num{0.5222601562825628} \\
        7 & \num{0.00002610518618947899} & \num{0.19325360225286092} & \colorbox{IRSColor}{\num{0.3425330670363406}} & \num{0.49045519958734163} \\
        8 & \num{0.00013358016007036788} & \num{0.16178593126662744} & \num{0.3392482545508067} & \num{0.4350950835292209} \\
        9 & \num{0.0015381992491483726} & \num{0.1982424097023041} & \num{0.26405408523341795} & \num{0.39552711582019207} \\
        10 & \num{0.00010511941192673934} & \num{0.14400035953845314} & \num{0.1796589524189155} & \num{0.29599279341113} \\
        11 & \num{0.000079478620551243} & \num{0.1312978717698955} & \num{0.1794530860920678} & \num{0.16836847891737716} \\
        12 & \num{0.00005261405457999838} & \colorbox{JEMMIGColor}{\num{0.1268995249833963}} & \num{0.18114335937291515} & \num{0.16273348181312874} \\
        13 & \num{0.000113467884525525} & \num{0.13767730728161975} & \num{0.17318770146334833} & \num{0.16126758397572893} \\
        14 & \num{0.000004656393674090089} & \num{0.1381282464868595} & \num{0.17641360568872672} & \num{0.16009127002685442} \\
        15 & \num{0.000019208659194882866} & \num{0.1355482340855555} & \num{0.17737483253173725} & \num{0.15677363978816228} \\
        All & \num{0.011202663939867546} & \num{0.2210467057016815} & \num{0.25657894824558186} & {\num{0.9895667535054589}} \\
        \bottomrule
    \end{tabular}
    \caption{Metrics Overview for Speaker Identification by Dimension. The top values for each metric are highlighted with cell colors including the ``All'' cases. We have highlighted the row that corresponds with the top-performant LP results, in this case, row 3.}
    \label{tab:speaker_id_metrics_overview}
\end{table}

With all scores normalized in the range of \([0,1]\), the results reveal relatively low MIG and IRS values for speaker identification, indicating that isolating this factor within single dimensions poses a challenge. In contrast, the combined Explicitness score shows improvement both dimension-wise and when aggregating all, suggesting a moderate ability of the model to represent speaker identity when multiple dimensions contribute. Notably, the Explicitness metric score of \(0.5796\) at index \(3\) correlates with the LP results, while IRS demonstrates the second-closest alignment regarding similar trends. Not surprisingly, among the metrics, the information-based scores exhibit the largest discrepancies from the LP assessment results this is primarily based on the deviations in the assumptions and algorithmic procedures used by both assessments where the former utilizes a non-linearity assumption while the latter a linearity relation. Also, intervention and predictor-based approaches use linear regressors or classifiers in the metric computation which is more or less a similar approach in LP analysis.

\subsubsection{Speaker Gender}

In table~\ref{tab:speaker_gender_metrics_overview}, we show the evaluation results for speaker gender across dimensions. Due to its binary nature, this task may be more solvable, allowing us to evaluate the model’s effectiveness in isolating it.

\begin{table}[!ht]
    \centering
    \begin{tabular}{@{}l S[table-format=1.6] S[table-format=1.6] S[table-format=1.6] S[table-format=1.6]@{}}
        \toprule
        Dimension & \multicolumn{2}{c}{\textbf{Information-based}} & \textbf{Intervention-based} & \textbf{Predictor-based} \\ 
        \cmidrule(r){2-3} \cmidrule(r){4-4} \cmidrule(r){5-5}
                  & {MIG $\uparrow$} & {JEMMIG $\downarrow$} & {IRS $\uparrow$} & {Explicitness $\uparrow$} \\ 
        \midrule
        0 & \num{0.00032222050363285417} & \num{0.318368} & \num{0.056089312218573065} & \num{0.28757517858651305} \\
        1 & \num{0.0008473753443538493} & \num{0.25702442836022266} & \num{0.04332216020439621} & \num{0.19028368877006763} \\
        2 & \num{0.0017888700554294854} & \num{0.364279} & \num{0.061134} & \num{0.226696} \\
        3 & \colorbox{MIGColor}{\num{0.003172846861139172}} & \num{0.34264613921323084} & \num{0.05844110959900089} & \num{0.220223} \\
        \rowcolor{LightGrayColor} 
        4 & \num{0.00019070201643377107} & \num{0.25981990557615664} & \num{0.068446} & \colorbox{ExplicitnessColor}{\num{0.586726}} \\
        5 & \num{0.00015122154432768318} & \num{0.3237202851428388} & \num{0.05857447726262034} & \num{0.2764552671428788} \\
        6 & \num{0.00021043536837796976} & \num{0.36661039646655325} & \num{0.062342229277334654} & \num{0.23583532076671876} \\
        7 & \num{0.00001375355294839023} & \num{0.353946} & \num{0.09540241525942605} & \num{0.379238} \\
        8 & \num{0.0000555212173384001} & \num{0.34686909399428767} & \colorbox{IRSColor}{\num{0.125838}} & \num{0.306610} \\
        9 & \num{0.00004814925320372875} & \num{0.290914} & \num{0.060501} & \num{0.412883} \\
        10 & \num{0.000024071119462015523} & \num{0.244427} & \num{0.037174} & \num{0.141321} \\
        11 & \num{0.00000139993724464869} & \colorbox{JEMMIGColor}{\num{0.232741}} & \num{0.037134} & \num{0.072569} \\
        12 & \num{0.000032624550826563226} & \num{0.237654} & \num{0.040399} & \num{0.071703} \\
        13 & \num{0.0000034404042955143922} & \num{0.260946} & \num{0.038271} & \num{0.066857} \\
        14 & \num{0.000007274158375215081} & \num{0.250887} & \num{0.036934} & \num{0.064348} \\
        15 & \num{0.000029817449599341524} & \num{0.247929} & \num{0.036131} & \num{0.061193} \\
        All & {\num{0.003172846861139172}} & \num{0.342646} & \num{0.059611} & {\num{0.870994}} \\
        \bottomrule
    \end{tabular}
    \caption{Metrics Overview for Speaker Gender by Dimension. The top values for each metric are highlighted with cell colors excluding the ``All'' row at the bottom, and the highlighted row $4$ indicates the corresponding best-performing LP index..}
    \label{tab:speaker_gender_metrics_overview}
\end{table}

For this result as well, all scores are also normalized on the same \([0,1]\) scale. The scores for this factor analysis exhibit higher JEMMIG and lower Explicitness values compared to other attributes, suggesting that gender may be more prone to be redundantly encoded across dimensions, as previously demonstrated in Figure~\ref{fig:accuracy_trends_humans_medium}, even though dimension \(4\) achieved the highest test accuracy. Additionally, we observe a minimal correlation between the highest performing LP-based dimension, \(4\), and the results shown in row \(4\) of Table~\ref{tab:speaker_gender_metrics_overview} except for the Explicitness score where we again see a direct correspondence as argued in the case of Speaker Identity factor analysis. However, the IRS score for dimension \(4\) aligns with the third-best LP result but has a minor deviation from the score for row $4$ of the present table.

\subsubsection{Speaking Style}
The assessment for speaking style, a fairly more complex attribute with multiple categories, is shown in Table~\ref{tab:speaking_styles_metrics_overview}. This evaluation reveals the model’s capacity to differentiate speaking styles.

\begin{table}[!ht]
    \centering
    \begin{tabular}{@{}l S[table-format=1.6] S[table-format=1.6] S[table-format=1.6] S[table-format=1.6]@{}}
        \toprule
        Dimension & \multicolumn{2}{c}{\textbf{Information-based}} & \textbf{Intervention-based} & \textbf{Predictor-based} \\ 
        \cmidrule(r){2-3} \cmidrule(r){4-4} \cmidrule(r){5-5}
                  & {MIG $\uparrow$} & {JEMMIG $\downarrow$} & {IRS $\uparrow$} & {Explicitness $\uparrow$} \\ 
        \midrule
            \rowcolor{LightGrayColor} 
            0 & \colorbox{MIGColor}{\num{0.008435}} & \num{0.214827} & \num{0.129198} & \num{0.472546} \\
            
            1 & \num{0.006420} & \num{0.199877} & \num{0.126353} & \colorbox{ExplicitnessColor} {\num{0.481304}} \\
            2 & \num{0.002108} & \num{0.280419} & \num{0.153095} & \num{0.404196} \\
            3 & \num{0.002197} & \num{0.229517} & \num{0.142767} & \num{0.383569} \\
            4 & \num{0.000181} & \num{0.227050} & \num{0.129853} & \num{0.304278} \\
            5 & \num{0.000518} & \num{0.243898} & \num{0.136215} & \num{0.395428} \\
            6 & \num{0.000115} & \num{0.236685} & \num{0.122339} & \num{0.322899} \\
            7 & \num{0.000118} & \num{0.279751} & \colorbox{IRSColor}{\num{0.161785}} & \num{0.280789} \\
            8 & \num{0.000408} & \num{0.333917} & \num{0.150182} & \num{0.201450} \\
            9 & \num{0.000513} & \num{0.233718} & \num{0.137986} & \num{0.204106} \\
            10 & \num{0.000416} & \num{0.195584} & \num{0.076734} & \num{0.211756} \\
            11 & \num{0.000044} & \num{0.204505} & \num{0.075087} & \num{0.105207} \\
            12 & \num{0.000033} & \num{0.196726} & \num{0.078986} & \num{0.078647} \\
            13 & \num{0.000003} & \colorbox{JEMMIGColor}{\num{0.190953}} & \num{0.072790} & \num{0.078846} \\
            14 & \num{0.000066} & \num{0.218774} & \num{0.073226} & \num{0.074521} \\
            15 & \num{0.000035} & \num{0.206115} & \num{0.075161} & \num{0.072820} \\
            All & \num{0.001889} & \num{0.197548} & \num{0.119298} & {\num{0.869577}} \\
        \bottomrule
    \end{tabular}
\caption{Metrics Overview for Speaking Style by Dimension. The top values for each metric are highlighted with cell colors sans the last row, and the highlighted row $0$ indicates the corresponding top-performing LP index.}
    \label{tab:speaking_styles_metrics_overview}
\end{table}

For this factor, MIG consistently shows lower scores, with especially small values in the later dimensions but the best score corresponds with the results of the LP analysis in Figure~\ref{fig:accuracy_trends_humans_medium}. We also observe a similar pattern for the second-best Explicitness score of $0.4725$. Unlike the previous two factors, we found the Explicitness scores on average, in this analysis, are lower which is a similar notion found in the LP case. Moreover, the JEMMIG results similarly indicate some correlation, though multiple dimensions—\(1\), \(10\), and \(15\)—are within close intervals. 
 
Disentanglement is a nuanced concept that, while intuitive to understand, is challenging to pinpoint within complex data scenarios, particularly those with temporal dynamics, such as audio. LP provides a simpler proxy for assessing it from learned representations, though it depends on performance metrics, training methods, and hyperparameter settings, among other considerations. Our results indicate some degree of factor disentanglement across key dimensions. 

Using the SDMs-based approach, we further assessed the modularity, compactness, and explicitness of the learned latent representations. We found that, as initially shown through LP, a degree of disentanglement is achievable, especially when examining isolated traits—such as Explicitness—rather than holistic methods. The correlation between the Explicitness score and LP results is intuitive, as both methods rely on a classifier or regressor to predict factors from latent representations.

Additionally, dimensions \(12\) through \(15\) appear to encode minimal information across factors for metrics like MIG, IRS, and Explicitness. JEMMIG displays a similar trend, except for the Speaker ID and Speaking Style factors. Notably, the Explicitness score for the aggregated dimensions (``All'') consistently achieves the best score across all three tables, indicating that combining dimensions may capture the relevant information more robustly which, however, comes at the price of not understanding which dimension makes what contribution. Additional results assessing moderate-to-low-level acoustic parameters are available in Appendix~\ref{appendix:results}, Tables~\ref{tab:rms_amplitude_metrics} and~\ref{tab:peak_amplitude_metrics}. Given that the learned latents are continuous, discretizing them for comparison with discrete factors may introduce discretization error, affecting the accuracy of information-based metrics.

\section{Conclusion}
\label{sec:conclusion}
In this study, we introduced SynSpeech, a large-scale synthetic speech dataset designed to benchmark disentangled representation learning in the speech domain. Our approach establishes a structured framework for assessing disentanglement by employing linear probing (LP) and supervised disentanglement metrics (SDMs) to evaluate representation learners like the RAVE model in terms of its effectiveness in isolating key generative factors such as speaker identity, gender, and speaking style, etc. Our results highlight the challenges of achieving comprehensive disentanglement for complex attributes like speaker identity, where factors remain partially entangled despite specialized encoding in select dimensions.

While LP provided insights into task-specific encoding, SDMs enabled us to systematically analyze modularity, compactness, and explicitness across the model’s latent dimensions. Notably, predictor-based SDMs naturally show more correlation with LP results while information theoretic shows the least agreement. 

This dataset and our evaluation framework offer a valuable resource for evaluating and comparing disentanglement techniques in speech processing. Future directions could include a more in-depth analysis of the sequential dependencies within each latent variable, exploration of representation transfer from synthetic to natural speech datasets, and application of transfer learning approaches. We anticipate that these contributions will support further research, encouraging both reproducibility and methodological innovation in disentangled representation learning for speech data. With continued refinement and exploration, disentangled representations may lead to advances in interpretable, fair, and adaptable speech models, enhancing performance across a broad range of speech-related applications.

\begin{ack}
This work was funds of the research training group (RTG) in ``Computational Cognition'' (GRK2340) provided by the Deutsche Forschungsgemeinschaft (DFG), Germany, and an EU-Consolidator grant (772000, TurnTaking).
\end{ack}
\bibliographystyle{plainnat}
\bibliography{references}
\appendix
\section{Appendix A: SynSpeech Dataset Supplementary Information}
\label{appendix:dataset}
Further details, including supplementary files with metadata like speaker demographics and speaking style distributions, are available \href{https://synspeech.github.io/#synspeech}{here}.

\section{Appendix B: Model Architecture and Training Details}
\label{appendix:models}

Here, we outline the key model configurations, training parameters, and hardware setup used in our experiments.

\textbf{Key Model and Training Configurations}:
\begin{itemize}
    \item \textbf{Sampling Rate (SR)}: The audio was resampled to a standardized sampling rate of $44,100$ Hz.
    \item \textbf{Sequence Length}: Each audio sequence had a fixed length of $256$ samples.
    \item \textbf{Latent Dimension}: The learned latent space has $128$, designed to capture disentangled attributes of speech.
    \item \textbf{SVD-Compact Latent Space}: After applying Singular Value Decomposition (SVD), the dimensionality of the latent space was reduced to \texttt{batch\_size} $\times$ 16 $\times$ \texttt{sequence\_length} (i.e., $\mathbb{R}^{128\times 16 \times 256}$), enhancing interpretability and compactness.
    \item \textbf{Channels and Signals}: Each audio signal was processed as a single channel, with a signal length of $131,072$ samples or a duration of $2.97$ seconds at $44.1$kHz SR.
\end{itemize}

\textbf{Training Parameters: Phase I and II}:
\begin{itemize}
    \item \textbf{Batch Size}: $8$
    \item \textbf{Number of Epochs}: $150$
    \item \textbf{Number of Workers}: 8 workers were used to parallelize data loading and preprocessing.
\end{itemize}

\textbf{Training Parameters: Linear Probing}:
\begin{itemize}
    \item \textbf{Batch Size}: $128$
    \item \textbf{Number of Epochs}: $100$
    \item \textbf{Number of Workers}: 8 workers were used to parallelize data loading and preprocessing.
\end{itemize}

\textbf{Hardware Configuration}:
The training and inference processes were conducted on NVIDIA RTX 8000 GPUs, utilizing the above configurations to ensure efficient processing of the SynSpeech dataset.

\section{Appendix C: Additional Results and Analysis}
\label{appendix:results}

This appendix includes supplementary analyses that extend the primary findings in the main paper.

\subsection{Spectrogram Comparison: Input, Reconstructed, and Generated Speech}
\begin{figure}[htbp]
    \centering
    \includegraphics[width=\textwidth]{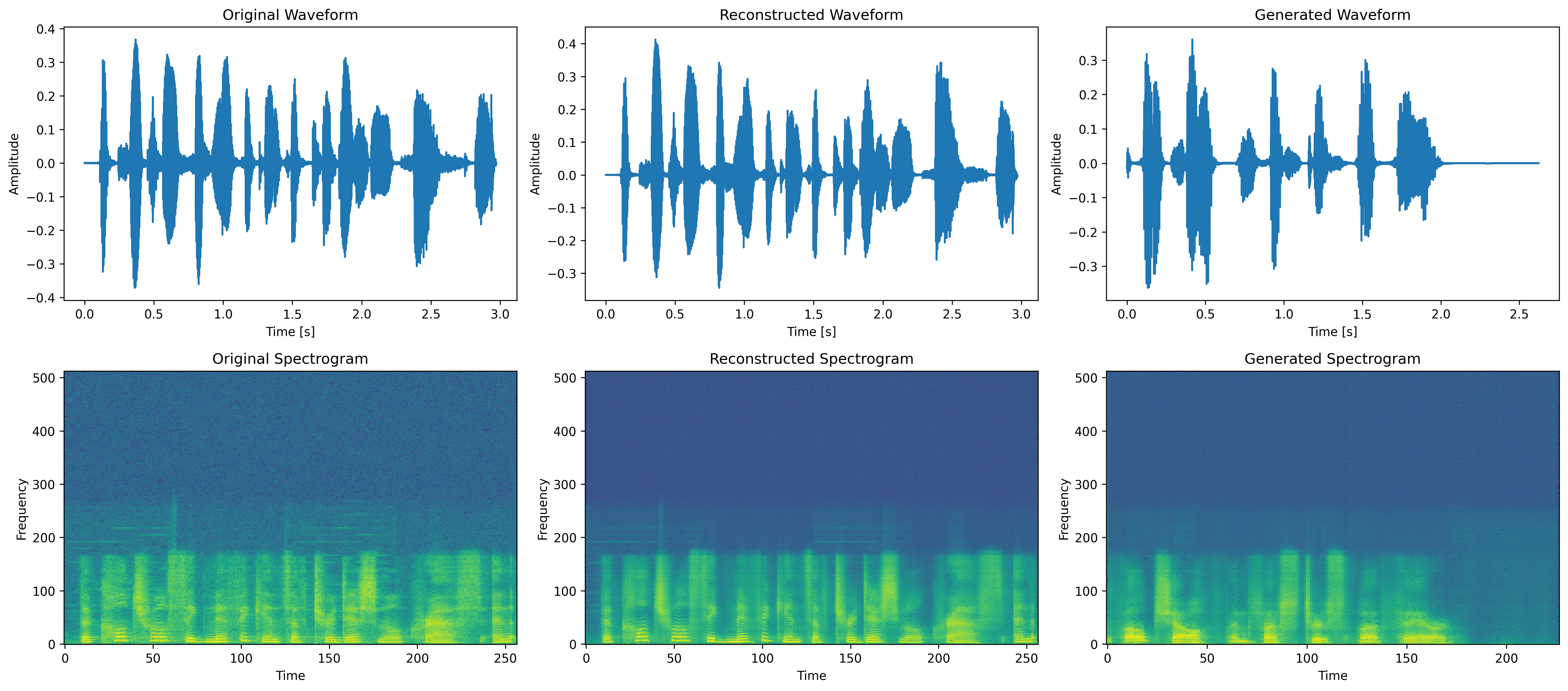}
    \caption{Comparison of original, reconstructed, and generated waveforms and spectrograms. The top row presents the waveform representations, illustrating the time-domain characteristics of the input, reconstructed, and generated audio signals. The bottom row shows the corresponding spectrograms, visualizing frequency content over time. The consistency between the original and reconstructed signals indicates effective preservation of temporal and spectral features, while the generated signal demonstrates the model’s ability to synthesize a plausible approximation of the original audio. Differences in the fine structure of the generated spectrogram suggest areas for improvement in capturing higher-frequency details and transient elements.}
    \label{fig:waveform_spec_batch0_sample0_waveform_spectrograms}
\end{figure}

Figure~\ref{fig:waveform_spec_batch0_sample0_waveform_spectrograms} provides a visual comparison of spectrograms for input, reconstructed, and generated speech samples, illustrating the model’s fidelity in capturing and reconstructing audio features with disentangled representations.

\subsection{Extended Linear Classifier Probing}
\begin{figure}[htbp]
    \centering
    \includegraphics[width=\textwidth]{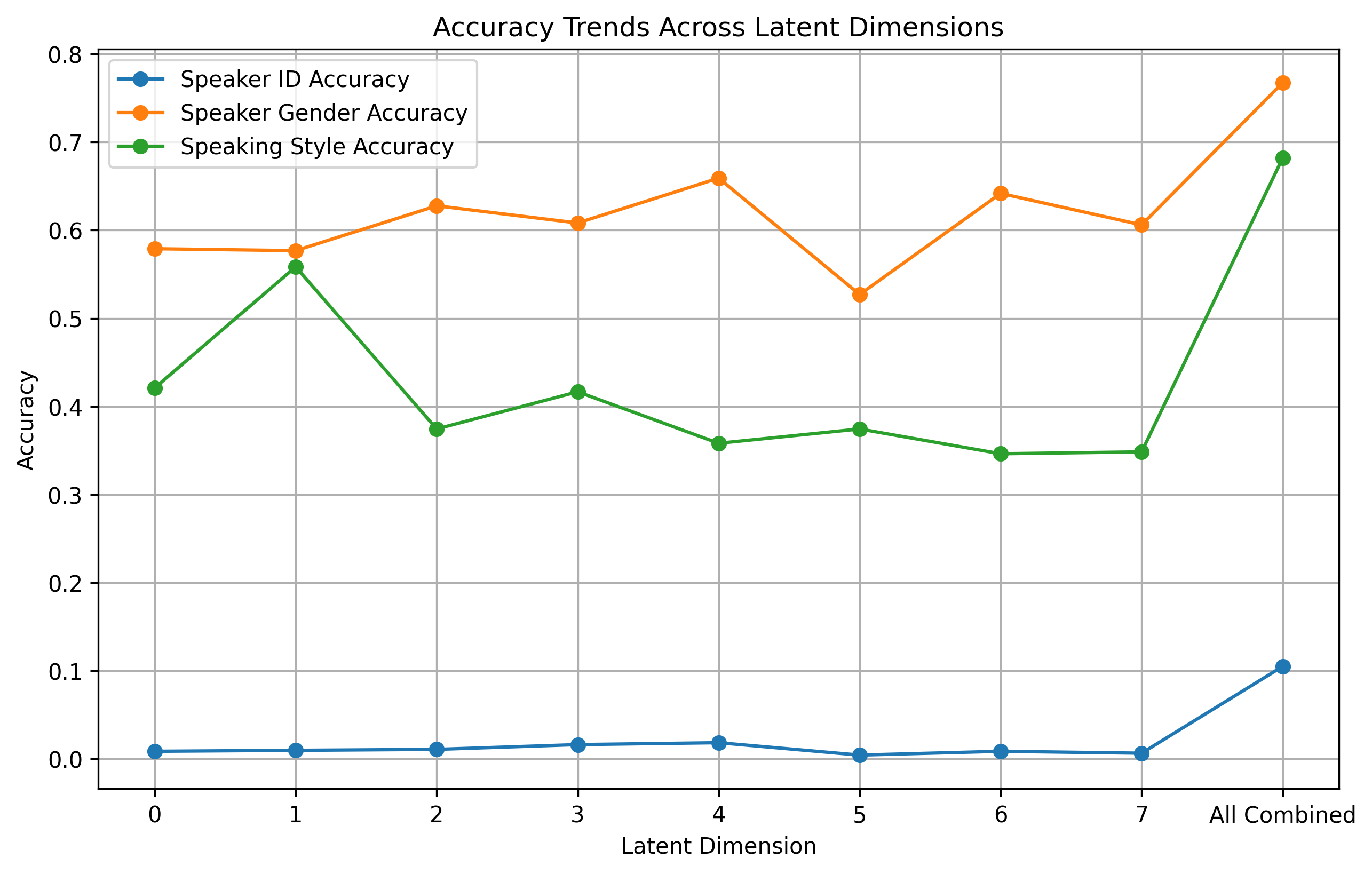}
    \caption{Linear probing accuracy trends across latent dimensions for the small version of the dataset, assessing the model's ability to predict speaker ID, speaker gender, and speaking style. The plot shows distinct accuracy patterns for each attribute, with speaker gender consistently achieving higher accuracy across dimensions, particularly in certain latent dimensions (e.g., 1 for speaking style and 4 for speaker gender), while speaker ID accuracy remains low throughout with some improvement on dimensions 3 and 4. The "All Combined" data point reflects the accuracy when using all latent dimensions together, highlighting an increase in accuracy across all attributes, especially for speaking style and speaker gender, suggesting that combined dimensions, unsurprisingly, capture richer representations.}
    \label{fig:accuracy_trends_humans_small}
\end{figure}
\subsection{Low-Level Acoustic Feature Disentanglement}
Tables~\ref{tab:rms_amplitude_metrics} and ~\ref{tab:peak_amplitude_metrics} present metrics assessing the disentanglement of low-level acoustic features, specifically RMS (Root Mean Square) Amplitude and Peak Amplitude. These features are crucial in audio signal processing, as they characterize different aspects of an audio signal’s intensity and dynamic range. RMS Amplitude represents the average power of the signal over a period, providing a measure of the signal's overall loudness or energy level. Peak Amplitude, on the other hand, captures the maximum instantaneous amplitude of the signal, highlighting the highest intensity points in the audio waveform. By analyzing these metrics, we can quantify the model’s ability to capture and isolate these specific acoustic attributes in a disentangled manner, which is essential for creating representations that retain meaningful, interpretable characteristics of the original audio. This evaluation helps determine if the model can effectively separate factors related to audio intensity from other features in the latent space.
\begin{table}[htb!]
    \centering
    \begin{tabular}{@{}l S[table-format=1.4] S[table-format=1.4] S[table-format=1.4] S[table-format=1.4]@{}}
        \toprule
        Dimension & \multicolumn{2}{c}{\textbf{Information-based}} & \textbf{Intervention-based} & \textbf{Predictor-based} \\ 
        \cmidrule(r){2-3} \cmidrule(r){4-4} \cmidrule(r){5-5}
                  & {MIG $\uparrow$} & {JEMMIG $\downarrow$} & {IRS $\uparrow$} & {Explicitness $\uparrow$} \\ 
        \midrule
            0 & \colorbox{MIGColor}{\num{0.003878}} & \num{0.273703} & \num{0.242911} & \colorbox{ExplicitnessColor} {\num{0.578852}} \\
            1 & \num{0.000250} & \colorbox{JEMMIGColor}{\num{0.173122}} & \num{0.263882} & \num{0.513448} \\
            2 & \num{0.000922} & \num{0.248553} & \num{0.265837} & \num{0.507579} \\
            3 & \num{0.000800} & \num{0.209992} & \num{0.274938} & \num{0.472308} \\
            4 & \num{0.000008} & \num{0.192558} & \num{0.235156} & \num{0.402866} \\
            5 & \num{0.000271} & \num{0.209368} & \num{0.270792} & \num{0.383198} \\
            6 & \num{0.000908} & \num{0.237163} & \num{0.263440} & \num{0.431474} \\
            7 & \num{0.002128} & \num{0.272765} & \colorbox{IRSColor}{\num{0.303663}} & \num{0.449524} \\
            8 & \num{0.000338} & \num{0.274407} & \num{0.294862} & \num{0.395023} \\
            9 & \num{0.000021} & \num{0.214034} & \num{0.258599} & \num{0.359220} \\
            10 & \num{0.000022} & \num{0.174427} & \num{0.211117} & \num{0.375654} \\
            11 & \num{0.000157} & \num{0.206873} & \num{0.210324} & \num{0.290618} \\
            12 & \num{0.000003} & \num{0.173165} & \num{0.212655} & \num{0.275420} \\
            13 & \num{0.000097} & \num{0.187901} & \num{0.206621} & \num{0.275356} \\
            14 & \num{0.000016} & \num{0.174974} & \num{0.207887} & \num{0.275878} \\
            15 & \num{0.000062} & \num{0.191128} & \num{0.209214} & \num{0.273917} \\
            All & \num{0.001533} & \num{0.248246} & \num{0.249539} & {\num{0.810331}} \\
        \bottomrule
    \end{tabular}
    \caption{Metrics by Dimension for the RMS Amplitude factor. The top values for each metric are highlighted with cell colors.}
    \label{tab:rms_amplitude_metrics}
\end{table}

\begin{table}[htb!]
    \centering
    \begin{tabular}{@{}l S[table-format=1.4] S[table-format=1.4] S[table-format=1.4] S[table-format=1.4]@{}}
        \toprule
        Dimension & \multicolumn{2}{c}{\textbf{Information-based}} & \textbf{Intervention-based} & \textbf{Predictor-based} \\ 
        \cmidrule(r){2-3} \cmidrule(r){4-4} \cmidrule(r){5-5}
                  & {MIG $\uparrow$} & {JEMMIG $\downarrow$} & {IRS $\uparrow$} & {Explicitness $\uparrow$} \\ 
        \midrule
            0 & \num{0.000232} & \num{0.270703} & \num{0.247637} & \num{0.428551} \\
            1 & \num{0.000740} & \colorbox{JEMMIGColor}{\num{0.176207}} & \num{0.258464} & \colorbox{ExplicitnessColor} {\num{0.465745}} \\
            2 & \colorbox{MIGColor}{\num{0.001876}} & \num{0.254614} & \num{0.289100} & \num{0.422461} \\
            3 & \num{0.000360} & \num{0.210067} & \num{0.284993} & \num{0.423584} \\
            4 & \num{0.000471} & \num{0.193776} & \num{0.266586} & \num{0.385694} \\
            5 & \num{0.000093} & \num{0.220893} & \num{0.262371} & \num{0.425096} \\
            6 & \num{0.000340} & \num{0.263004} & \num{0.269600} & \num{0.414080} \\
            7 & \num{0.000734} & \num{0.275145} & \colorbox{IRSColor}{\num{0.304745}} & \num{0.387866} \\
            8 & \num{0.000399} & \num{0.277735} & \num{0.287724} & \num{0.341978} \\
            9 & \num{0.000440} & \num{0.224564} & \num{0.262137} & \num{0.343523} \\
            10 & \num{0.000100} & \num{0.188947} & \num{0.212561} & \num{0.314930} \\
            11 & \num{0.000021} & \num{0.212072} & \num{0.209169} & \num{0.294456} \\
            12 & \num{0.000020} & \num{0.178009} & \num{0.212925} & \num{0.275805} \\
            13 & \num{0.000027} & \num{0.188516} & \num{0.207636} & \num{0.283400} \\
            14 & \num{0.000013} & \num{0.182404} & \num{0.210053} & \num{0.286994} \\
            15 & \num{0.000001} & \num{0.183930} & \num{0.211195} & \num{0.292490} \\
            All & \num{0.000740} & {\num{0.176207}} & \num{0.253898} & {\num{0.735209}} \\
        \bottomrule
    \end{tabular}
    \caption{Metrics by Dimension for the  Peak Amplitude factor. The top values for each metric are highlighted with cell colors.}
    \label{tab:peak_amplitude_metrics}
\end{table}




\end{document}